\documentclass[aps,prl,twocolumn,amsmath,amssymb,showpacs,superscriptaddress]{revtex4-2}
\usepackage{graphicx}
\usepackage{amssymb}
\usepackage{bm}
\usepackage{braket}
\usepackage{amsmath}
\usepackage{tabularx}
\usepackage{ltablex}
\usepackage{multirow}
\usepackage[dvipsnames]{xcolor}

\usepackage[colorlinks=true,
                linkcolor=blue,
	        citecolor=blue,
	        urlcolor=blue]{hyperref}

\newcommand{\trieste}{Dipartimento di Fisica, Universit\`a di Trieste, Strada Costiera 11, I-34151 Trieste, Italy}

\usepackage{verbatim}

\newcommand\code[1]{\texttt{#1}}
\AtBeginDocument{\usepackage{booktabs}}
\begin{document}
\title{Local Chern Marker for Periodic Systems}
\date{\today}
\author{Nicolas Baù}
\affiliation{\trieste}
\author{Antimo Marrazzo}
\email{antimo.marrazzo@units.it}
\affiliation{\trieste}

\begin{abstract}
Topological invariants are global properties of the ground-state wave function, typically defined as winding numbers in reciprocal space. Over the years, a number of topological markers in real space have been introduced, allowing to map topological order in heterogeneous crystalline and disordered systems. 
Notably, even if these formulations can be expressed in terms of lattice-periodic quantities, they can actually be deployed in open boundary conditions only, as in practice they require computing the position operator $\mathbf{r}$ in a form that is ill-defined in periodic boundary conditions. 
Here we derive a local Chern marker for infinite two-dimensional systems with periodic boundary conditions in the large supercell limit, where the electronic structure is sampled with one single point in reciprocal space. We validate our approach with tight-binding numerical simulations on the Haldane model, including trivial/topological superlattices made of pristine and disordered Chern insulators. The strategy introduced here is very general and could be applied to other topological invariants and quantum-geometrical quantities in any dimension. 
\end{abstract}
\maketitle
Topological order is defined by the existence of integer topological invariants that globally characterize the system and cannot be changed under adiabatic transformations. For condensed-matter systems, and electronic structure in particular, topological invariants are often introduced as reciprocal-space integrals of quantum-geometrical quantities that become quantized over closed surfaces. The archetypal topological invariant is the Chern number in two dimensions, which can be written as the integral of the Berry curvature over the Brillouin zone (BZ) torus and it is a $\mathbb{Z}$ integer~\cite{vanderbilt_book_2018,bernevig_book_2013}. Even for correlated systems, integer invariants can be obtained by integrating over the boundary conditions, as for the many-body Chern number~\cite{thouless_prb_1985}. Hence, topological invariants are conventionally introduced as global quantities of the system, where the latter is implicitly assumed to be homogeneous. A number of strategies have been proposed to calculate topological invariants in non-homogeneous or disordered systems~\cite{corbae_EPL_2023}, such as the switch-function formalism~\cite{Elbau2002, PhysRevB.97.195312, 10.1063/5.0096720}, methods based on the scattering matrix~\cite{PhysRevB.85.165409, PhysRevB.83.155429} or on the non-commutative index theorem~\cite{Avron1994, AVRON1994220, 10.1063/1.5026964}, the Bott index~\cite{10.1063/1.3274817, HASTINGS20111699, Loring_2011}, real-space formulas~\cite{prodan_prl_2010,prodan_NJP_2010} and local markers~\cite{bianco_prb_2011}.

Notably, non-trivial topology has its most dramatic manifestation in the simplest, and most relevant, non-homogeneous setting: the interface between a topological phase and a trivial phase, such as vacuum or a trivial semiconductor. For the sake of concreteness, let us consider a quantum anomalous Hall (a.k.a. Chern) insulator in absence of strong electronic correlations, which is defined by a non-vanishing Chern number, and we place it in contact with vacuum or with a trivial insulator. As topological invariants cannot change under deformations of the Hamiltonian that do not close the band gap, metallic chiral edge states will appear at the one-dimensional edge separating the trivial from the topological regions. This setting cannot be described at once by any global topological invariant, as the system is clearly made of two, topologically distinct, regions. On the other hand, the locality (or ``nearsightedness'' according to Kohn~\cite{kohn_prl_1996}) of the ground-state electronic structure hints that it must be possible to probe non-trivial topology, which is also a ground state property, locally in real space and by using only knowledge from the neighborhood around the region of interest.

Indeed, it has been shown by Bianco and Resta~\cite{bianco_prb_2011} that the Chern number $C$ can be mapped to real space by deriving a topological marker $\mathfrak{C}(\mathbf{r})$. Notably, this is possible not only for topological\textemdash i.e., quantized\textemdash properties of the electronic structure, but also for the broader set of quantum-geometrical quantities, where a corresponding local real-space marker exists for both insulators and metals. Examples include the geometrical intrinsic part of the anomalous Hall conductivity~\cite{marrazzo_prb_2017,rauch_prb_2018}, the orbital magnetization~\cite{bianco_prl_2013,marrazzo_prl_2016,bianco_prb_2016,dhv_prb_2013} and the localization tensor~\cite{marrazzo_prl_2019} which is deeply connected to the quantum metric~\cite{provost_cmp_1980}. Local topological markers can be related to local circular dichroism, which can be experimentally measured~\cite{Pozo_PRL_2019,Marsal_PNAS_2020,Molignini_scipost_2023}.
 The strategy behind the Bianco-Resta marker relies on rewriting the Berry curvature in terms of the operator $\mathcal{P}\mathbf{r}\mathcal{Q}$, where $\mathcal{P}$ and $\mathcal{Q}$ are the projector operators over the occupied and empty orbitals respectively. Several equivalent formulations of the local Chern marker (LCM) are possible, here we report the one developed in Ref.~\cite{marrazzo_prb_2017}:
\begin{eqnarray}
  \label{eq:lcm1} \mathfrak{C}(\mathbf{r}) &=& -4\pi\mathrm{Im}  \braket{\mathbf{r}|\mathcal{P}x\mathcal{Q}y|\mathbf{r}}\\
&=& 4\pi\text{Im} \braket{\mathbf{r}|\mathcal{P}\left[x,\mathcal{P}\right]\left[y,\mathcal{P}\right]|\mathbf{r}},\label{eq:lcm2}
\end{eqnarray}
where $x$ and $y$ are the Cartesian components of the position operator $\mathbf{r}$; the second line is reformulated to emphasize that only occupied states are needed, hence being more suited to numerical implementations. 
The local Chern number can then be obtained by calculating the macroscopic average of $\mathfrak{C}(\mathbf{r})$, essentially taking the trace per unit of area $\mathrm{Tr}_A$ in a neighborhood around the region of interest.
A key aspect of this approach is that the Chern number can be expressed as a \emph{trace}, which can be evaluated locally in real space. At variance with $\mathbf{r}$, the operators $\mathcal{P}\mathbf{r}\mathcal{Q}$ and  $[\mathbf{r},\mathcal{P}]$ are well-defined and regular even in periodic boundary conditions (PBCs)~\cite{bianco_prb_2011}. However, in practice, such operators are not directly accessibile and are typically constructed by calculating first the position operator $\mathbf{r}$ in a proper basis: only after, $\mathbf{r}$ is multiplied with the projection operators to obtain Eqs.~\ref{eq:lcm1} or \ref{eq:lcm2}. Hence, the usual LCM can only be applied to open boundary conditions (OBCs), i.e., for finite samples. 

In this work, we derive a novel LCM for extended systems in PBCs. Our approach allows calculating space-resolved Chern numbers from large-cell electronic structure simulations, where the BZ is typically sampled with a single point, usually the BZ center $\Gamma$. We validate the approach on homogeneous samples and trivial/topological superlattices based on the Haldane model~\cite{haldane_1988}, where we examine both pristine and disordered Chern insulating phases. In passing, we provide a simple physical picture that connects the single-point Chern number~\cite{ceresoli_sp_prb_2007,favata_es_2023} with the Bott index~\cite{10.1063/1.3274817, HASTINGS20111699, Loring_2011}, and demonstrate that both methods measure the topological obstruction to choose a periodic smooth gauge all over the BZ in the large-cell (i.e., single $k$-point) limit.

PBCs are often the method of choice for electronic structure calculations. Not only they are the most natural option to study perfect crystals, but they also reduce finite size-effects for non-crystalline systems, such as materials with defects, surfaces, heterojunctions (e.g., in lead-conductor-lead geometries) or amorphous materials. In the case of non-crystalline structures, large periodic cells are used and results are typically checked for convergence with respect to the size of the simulation cell. If the size is large enough, the sampling of the BZ can be reduced to a single point in reciprocal space, typically the $\Gamma$ point. In this limit, the Chern number can be calculated through single-point formulas as discussed in Ref.~\cite{ceresoli_sp_prb_2007}, we report here the result derived therein:
\begin{eqnarray}\label{eq:pbclcm_1}
    C=-\frac{1}{\pi}\mathrm{Im}\sum_{n=1}^{N_{occ}} \langle\tilde u_{n\mathbf b_1}|\tilde u_{n \mathbf b_2}\rangle,
\end{eqnarray}
where $\mathbf{b}_{1,2}$ are the reciprocal lattice vectors, $|\tilde u_{n\mathbf b_j}\rangle$ are the ``dual'' states of the Hamiltonian eigenstates $|u_{n\Gamma}\rangle$, and represent, in the limit of a large supercell, the states obtained by parallel transport.
The sum runs over $N_{occ}$ occupied dual states defined as:
\begin{equation}
  \label{eq:duals_def}
  \ket{\tilde u_{n\mathbf b_j}}=\sum_{m=1}^{N_{occ}}S^{-1}_{mn}(\mathbf b_j)e^{-i\mathbf b_j\cdot\mathbf r}\ket{u_{m\Gamma}}
\end{equation}
where we introduce the overlap matrix $S_{nm}(\mathbf b_j) = \braket{u_{n\Gamma}|e^{-i\mathbf b_j\cdot\mathbf r}|u_{m\Gamma}}$. The dual states enjoy the property $\langle\tilde u_{n\mathbf b_j}|u_{m\Gamma}\rangle=\delta_{nm}$ and allow fixing a continuous gauge, essentially adopting a discretized version of the covariant derivative~\cite{sai_prb_2002,souza_prb_2004}.

Now we rewrite Eq.~\ref{eq:pbclcm_1} as a trace
\begin{equation}
  \label{eq:pbclcm_2}
  C=-\frac{1}{\pi} \mathrm{Im} \mathrm{Tr} \left\{\sum_{n=1}^{N_{occ}} \ket{\tilde u_{n\mathbf b_2}}\bra{\tilde u_{n\mathbf b_1}}\right\}
\end{equation}
and we define
\begin{eqnarray}
  \label{eq:def_p}
 \tilde P_{\mathbf b_2,\mathbf b_1}=\sum_{n=1}^{N_{occ}}|\tilde u_{n\mathbf b_2}\rangle\langle \tilde u_{n\mathbf b_1}|.
\end{eqnarray}
In general $\tilde P_{\mathbf b_2,\mathbf b_1}$ is not a projector but can be written in terms of the projectors $\mathcal P_{\mathbf b_j}=\sum_{n=1}^{N_{occ}}\ket{\tilde u_{n\mathbf b_j}}\bra{\tilde u_{n\mathbf b_j}}$:
\begin{eqnarray}
  \label{eq:prod_proj}
	\tilde P_{\mathbf b_2,\mathbf b_1} = \mathcal{P}_{\mathbf b_2}\mathcal{P}_{\Gamma}\mathcal{P}_{\mathbf b_1}
\end{eqnarray}
where $\mathcal P_{\Gamma}=\sum_{n=1}^{N_{occ}} \ket{u_{n\Gamma}}\bra{u_{n\Gamma}}$. We exploit the cyclic property of the trace and write the single-point Chern number as
\begin{equation}
  \label{eq:sp_proj_comm}
  C = -\frac{1}{2\pi} \mathrm{Im} \mathrm{Tr} \left\{ \big[ \mathcal P_{\mathbf b_1}, \mathcal P_{\mathbf b_2} \big] \mathcal P_{\Gamma} \right\}.
\end{equation}
 Finally, the LCM can be evaluated by taking the macroscopic average (i.e., the local trace per unit of area) of
\begin{eqnarray}\label{eq:pbclcm_3}
	\mathcal{C}(\mathbf r)=-\frac{1}{2\pi}\mathrm{Im}\braket{\mathbf{r}|\big[ \mathcal P_{\mathbf b_1}, \mathcal P_{\mathbf b_2} \big] \mathcal P_{\Gamma} |\mathbf{r}}.
\end{eqnarray}
Eq.~\ref{eq:pbclcm_3} is manifestly gauge invariant and perspicuous: a non-vanishing local Chern number arises when $\mathcal{P}_{\mathbf b_1}$ and $\mathcal{P}_{\mathbf b_2}$ \emph{locally} do not commute. If fact, only in the trivial phase it is possible to choose a periodic smooth gauge such that $\mathcal{P}_{\mathbf b_1}=\mathcal{P}_{\mathbf b_2}$. 

Notably, this brings some insights on the common approach of calculating the Chern number of non-crystalline systems through the Bott index~\cite{loring_1991,bellissard_1994,huang_prl_2018,huang_prb_2018,Loring_2011,Toniolo_2022}, which measures the commutativity of projected position operators $U=\mathcal{P}_{\Gamma}e^{\frac{i2\pi}{L}x}\mathcal{P}_{\Gamma}$ and $V=\mathcal{P}_{\Gamma}e^{\frac{i2\pi}{L}y}\mathcal{P}_{\Gamma}$. Indeed, if we take the $\mathcal P_{\mathbf b_1}$ and $\mathcal P_{\mathbf b_2}$ operators as defined above Eq.~\ref{eq:prod_proj} but adopt a periodic gauge, we obtain the $U$ and $V$ operators appearing in the Bott index, as by definition of direct and reciprocal lattice vectors $\mathbf{a}_i\cdot\mathbf{b}_j=2\pi\delta_{ij}$. While Toniolo has mathematically demonstrated the formal equivalence between the Bott index and the Chern number~\cite{Toniolo_2022,Toniolo_prb_2018}, Eqs.~\ref{eq:sp_proj_comm} and~\ref{eq:pbclcm_3} provide a more physical explanation on why the non-commutativity of the projected position operators is related to the Chern number of non-crystalline systems: these $U$ and $V$ operators are essentially the ground-state projector transported from the BZ center to the BZ boundary with the periodic gauge.  As said above for the $\mathcal{P}_{\mathbf b_1},\mathcal{P}_{\mathbf b_2}$ operators, in presence of non-vanishing Chern numbers the gauge cannot be chosen simultaneously periodic and smooth over all the BZ, hence the two operators cannot commute. Hence, the single-point Chern number and the Bott index ultimately measure the same phenomenon: the topological obstruction to choose a periodic smooth gauge all over the BZ in the limit where the BZ shrinks to a single point.

In addition, Ref.~\cite{huang_prb_2018} reports that performing singular value decomposition (SVD) on the projected position operators improves the numerical stability of the Bott index calculation and is hence adopted in their study; no theoretical justification was provided therein, but from the point of view of Eq.~\ref{eq:sp_proj_comm} that can be understood as a way to invert the overlap matrix appearing in Eq.~\ref{eq:duals_def} and perform a single-point covariant derivative. However, we emphasize that the formula for the Bott index (see Eq. 4 of Ref.~\cite{huang_prb_2018}) still differs substantially from our PBCs LCM introduced in Eq.~\ref{eq:pbclcm_3}, including the fact that the Bott index calculates a logarithm and does not lead to a space-resolved local marker. 

We now consider the limit of an infinitely large supercell in PBC where a finite sample is surrounded by vacuum, so $x,y\ll L$ whenever the wavefunction is not zero and the sample is centered in the middle of a square cell. We can expand the projectors up to order $L^{-2}$ in the supercell size
\begin{equation}
\mathcal{P}_{\mathbf{b}_j} \approx \mathcal{P}_{\Gamma} + \frac{2\pi i}{L} \left[ \mathcal{P}_{\Gamma},r_j\right]-\frac{2\pi^2}{L^2}\left[r_j,\left[r_j,\mathcal P_{\Gamma}\right]\right]
\end{equation}
and use the expression in Eq.~\ref{eq:pbclcm_3}. Our PBC Chern marker essentially converges to the Bianco-Resta OBC Chern marker of Eq.~\ref{eq:lcm2} (full calculation in the Supplementary Material~\cite{SM}):
\begin{align}
  \mathcal{C}(\mathbf r)&\underset{x,y\ll L}= - 4\pi \text{Im}\bra{\mathbf r} \mathcal P_{\Gamma}x(\mathbb I-\mathcal P_{\Gamma})y \ket{\mathbf r} + \mathcal{O}(L^{-3}).
\end{align}

Before discussing numerical results, we introduce a ``symmetric'' version of our PBC marker, where we start from the corresponding ``symmetric'' single-point Chern number
\begin{align}
C^{(sym)}=\frac{-1}{4\pi}\mathrm{Im}\sum_{n=1}^{N_{occ}}&\left(\bra{\tilde u_{n\mathbf b_1}}-\bra{\tilde u_{n-\mathbf b_1}}\right)\left(\ket{\tilde u_{n\mathbf b_2}}-\ket{\tilde u_{n-\mathbf b_2}}\right),
\end{align}
and define the operator
\begin{align}
	&\bar P = \sum_{\sigma_1=\pm}\sum_{\sigma_2=\pm}\sigma_1\sigma_2\tilde P_{\sigma_2\mathbf b_2,\sigma_1\mathbf b_1}= \nonumber \\
	&=\frac{1}{2} \left(\, \sum_{\sigma_1=\pm}\sum_{\sigma_2=\pm}\sigma_1\sigma_2\big[\mathcal P_{\sigma\mathbf b_1},\mathcal P_{\sigma_2\mathbf b_2}\big]\right) \mathcal P,
\end{align}
where we generalize Eq.~\ref{eq:def_p} as
\begin{eqnarray}
 \tilde P_{\pm\mathbf b_2,\pm\mathbf b_1}=\sum_{n=1}^{N_{occ}}|\tilde u_{n\pm\mathbf b_2}\rangle\langle \tilde u_{n\pm\mathbf b_1}|.
\end{eqnarray}
Here the term ``symmetric'' refers to the symmetric derivative (as opposed to the right-hand, i.e., ``asymmetric'', derivative) used to calculate the Berry curvature when the single-point limit is taken~\cite{ceresoli_sp_prb_2007}. By performing similar steps as done for Eq.~\ref{eq:pbclcm_3}, one obtains a symmetric marker:
\newline
\begin{eqnarray}
	C^{(sym)}(\mathbf r)&=&-\frac{1}{4\pi}\mathrm{Im}\braket{\mathbf{r}| \bar{P} |\mathbf{r}} \nonumber \\
  &=&-\frac{1}{8\pi}\mathrm{Im}\bra{\mathbf{r}} \left(\left[\mathcal{P}_{\mathbf{b}_1},\mathcal{P}_{\mathbf{b}_2}\right]+\left[\mathcal{P}_{-\mathbf{b}_1},\mathcal{P}_{-\mathbf{b}_2}\right] \right. \nonumber  \\
  &&  - \left.\left[\mathcal{P}_{-\mathbf{b}_1},\mathcal{P}_{\mathbf{b}_2}\right]-\left[\mathcal{P}_{\mathbf{b}_1},\mathcal{P}_{-\mathbf{b}_2}\right]\right)\mathcal{P}_{\Gamma} \ket{\mathbf{r}}.\label{eq:pbc_lcm_asym}
\end{eqnarray}

We validate our approach with tight-binding simulations on the Haldane model~\cite{haldane_1988}, which describes electrons hopping on a 2D honeycomb lattice with a staggered magnetic flux. The parameters of the model are the nearest-neighbor hopping $t=1$, the on-site energy term $\pm\Delta$ with opposite signs on the two sublattices and the second-nearest-neighbor hopping term $t_2e^{i\phi}$ where $t_2=0.15$. We release~\footnote{The calculation of the local Chern marker in periodic boundary conditions is available in version 0.3.0 of the \code{StraWBerryPy} code package \url{https://github.com/strawberrypy-developers/strawberrypy/releases/tag/v0.3.0}.} a Python implementation of the LCM PBC in the \code{StraWBerryPy}~\cite{strawberrypy} package, which is interfaced with two popular tight-binding software engines such as \code{PythTB}~\cite{pythtb} and \code{TBmodels}~\cite{TB,tbmodels}, and it can be easily interfaced to other codes. 

In Fig.~\ref{fig:1}, we report the convergence of our PBC LCM calculated in a single cell (two sites) in the middle of the bulk, in its symmetric (Eq.~\ref{eq:pbc_lcm_asym}) and asymmetric implementation (Eq.~\ref{eq:pbclcm_3}), with respect to the supercell size. We also compare it with the Bianco-Resta Chern marker~\cite{bianco_prb_2011} calculated on finite samples within OBC corresponding to the same number of sites.
The convergence of the PBC LCM is described by a power law: for this choice of the Hamiltonian parameters the power is about $-3.6$ for the topological and about  $-2.7$ for the trivial phase if evaluated via the symmetric formula (Eq.~\ref{eq:pbc_lcm_asym}), while values are about half in the asymmetric version (Eq.~\ref{eq:pbclcm_3}). The value of the power depends on the magnitude of the gap, with larger gaps leading to faster convergence; in general small lattice sizes are sufficient to infer the topological phase with accuracy. As for the single-point Chern number~\cite{ceresoli_sp_prb_2007,favata_es_2023}, the symmetric formula (Eq.~\ref{eq:pbc_lcm_asym}) converges much faster than the asymmetric version (Eq.~\ref{eq:pbclcm_3}), hence in the following we always use Eq.~\ref{eq:pbc_lcm_asym}.
\begin{figure}[h!]
  \centering
  \includegraphics[width=1\linewidth]{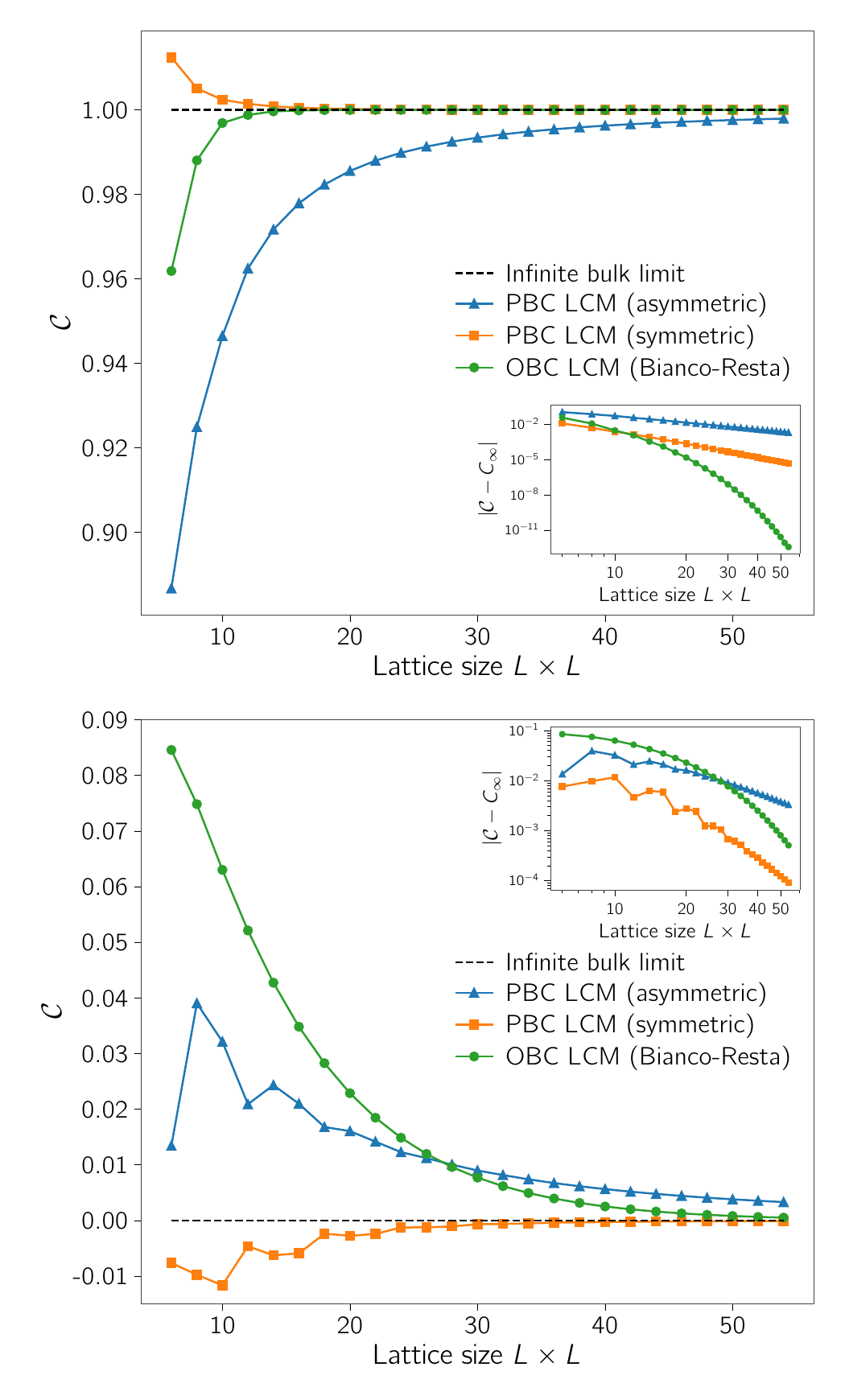}
  \caption{Convergence of the local Chern marker in periodic boundary conditions for the topological and trivial phase, comparison with the Bianco-Resta marker in open boundary conditions. Both the asymmetric and symmetric formulas converge with the supercell linear size $L$ as a power law, the symmetric version is notably faster.}
  \label{fig:1}
\end{figure}

Then we consider the more challenging case of a topological/trivial superlattice, where the system is perfectly periodic along one direction while it is made of alternating topological ($\Delta=0.3$) and trivial ($\Delta=1.25$) regions along the other. The results for a $6000$-site supercell are reported in Fig.~\ref{fig:2}. The PBC LCM neatly charts the topological landscape in real space and distinguishes the topological from the trivial region, even close to the boundary of the cell where Eqs.~\ref{eq:lcm1}-\ref{eq:lcm2} would dramatically fail. The left and right boundary of the cell are connected by PBC and the marker is continuous; topologically protected metallic 1D channels are present at about $1/3$ and $2/3$ of the cell along the direction $x$, precisely where the topological marker changes value. 
\begin{figure*}[ht]
  \centering
  \includegraphics[width=0.95\linewidth]{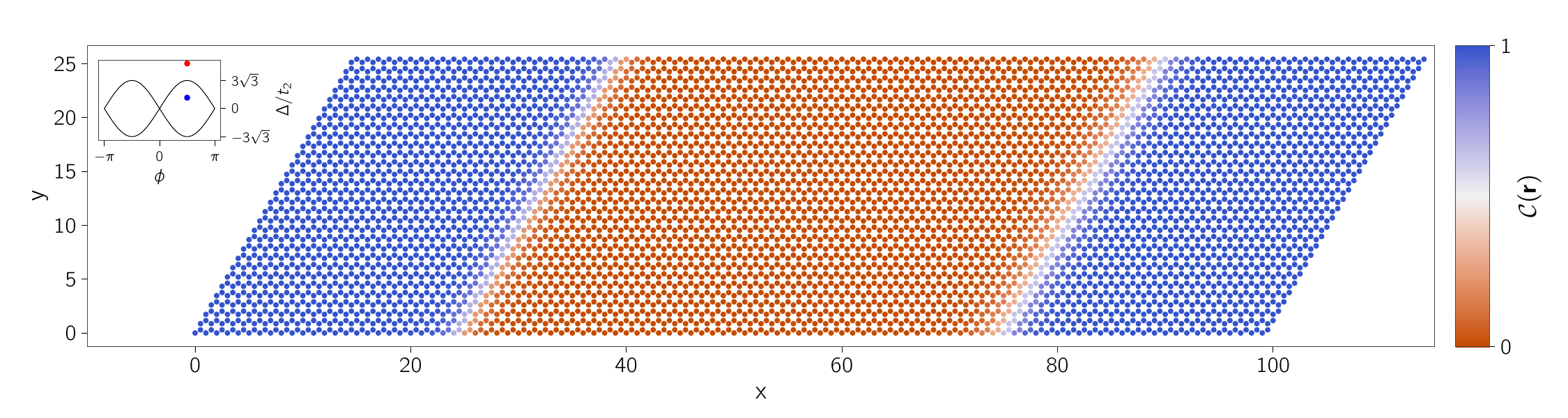}
  \caption{Local Chern marker for a $6000$-site superlattice of the pristine Haldane model made of topological and trivial stripes in periodic boundary conditions. The left and right regions are topological (Chern number $C=1$) while the center is trivial ($C=0$), one-dimensional metallic edge states separate the regions with different Chern numbers. The inset displays the model parameters used for the trivial (red) and topological (blue) regions.}
  \label{fig:2}
\end{figure*}
\begin{figure}[ht]
  \centering
  \includegraphics[width=1\linewidth]{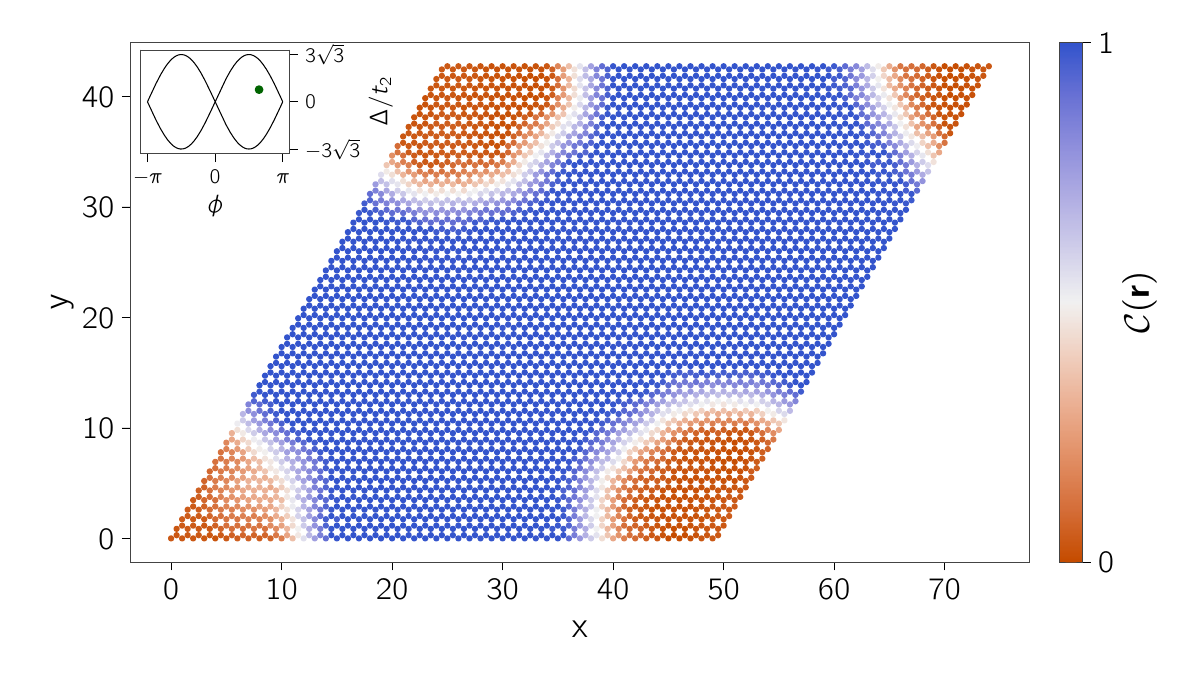}
  \caption{Local Chern marker for a $5000$-site supercell of the Haldane model with Anderson disorder $W$ in periodic boundary conditions. The system is made of disks of radius $R=13$ in the trivial phase (Chern number $C=0$) centered on the Bravais lattice, embedded in a topologically non-trivial matrix ($C=1$); circular metallic edge states separate the two regions with different Chern numbers. The parameters of the Haldane model are constant everywhere ($\Delta=0.2$, $\phi = 0.65\pi$), while the topological and trivial regions experience respectively weak ($W=1$) and strong ($W=10$) Anderson disorder. Smearing temperature is set to $T=0.05$ and macroscopic averages are performed over a circular region of radius $r=3$.}
  \label{fig:3}
\end{figure}

Last, we validate our approach for a non-homogenous \emph{and} disordered system. We consider a $5000$-site periodic supercell of the Haldane model ($\Delta=0.2$, $\phi = 0.65\pi$) made of disks of radius $R=13$ (in units of the lattice parameter) centered on the Bravais lattice with strong ($W=10$) Anderson disorder~\cite{abrahams_prl_1979, PhysRevB.104.214207}, embedded in a weakly-disordered ($W=1$) matrix. Disorder is introduced through a random on-site term uniformly distributed in $\left[-W/2,W/2\right]$. 
From single-point Chern number calculations we know that extended systems described by those Hamiltonian parameters would be respectively in the topological phase ($C=1$) for weak disorder and in the trivial phase ($C=0$) for strong disorder: hence, we expect our inhomogeneous system to display topologically trivial islands, embedded in a topologically non-trivial matrix.
This setting models topological insulators where the concentration of impurities or defects is not homogeneous (local damage), and locally sufficiently higher to drive the system to the trivial phase. The presence of disorder and metallic interfaces may lead to gapless density of states over large areas, so we adopt the usual smearing technique similarly to what has been done in Ref.~\cite{marrazzo_prb_2017} to study the locality of the anomalous Hall conductivity and improve the convergence with respect to the system size. In presence of smearing, the projectors $\mathcal P_{\mathbf b_j}$ sum over all the Hamiltonian eigenstates with a Fermi-Dirac occupation function $f(\epsilon_n, T, \mu)$, where $\epsilon_n$ is the eigenvalue corresponding to the $n$-th eigenstate, $T$ is a fictitious temperature and $\mu$ is the chemical potential. In addition, we introduce an upper cutoff by discarding the empty states with small occupations (for which $f(\epsilon_n, T,\mu)<f_{c}$ where we set $f_c=0.1$), so that the projectors appearing in Eqs.~\ref{eq:pbclcm_3} and \ref{eq:pbc_lcm_asym} take the form
\begin{align}
  &\mathcal P_{\Gamma} = \sum_{n:f(\epsilon_n,T,\mu)<f_c} f(\epsilon_n, T, \mu)\ket{ u_{n\Gamma}}\bra{ u_{n\Gamma}}, \\
  &\mathcal P_{\mathbf b_j} =\sum_{n:f(\epsilon_n,T,\mu)<f_c} f(\epsilon_n, T,\mu)\ket{\tilde u_{n\mathbf b_j}}\bra{\tilde u_{n\mathbf b_j}}.
\end{align}

We report in Fig.~\ref{fig:3} the numerical results: the PBC LCM is equal to one in the low-disorder regions and vanishing in the strong-disorder circular regions, where Anderson disorder is sufficiently strong to locally destroy the non-trivial topology. As the trivial islands are centered on the Bravais lattice, the PBC marker correctly displays them at the four corners of the supercell (a task that could not be achieved with the Bianco-Resta OBC marker), hence demonstrating that it correctly treats the position operator in PBCs while truly being a local marker. 

In conclusion, we have shown that topological order can be probed locally also in PBCs by means of a simple space-resolved marker, that is capable to chart electronic topology in inhomogeneous and/or disordered systems. Our PBC LCM is based on the ground-state electron distribution only and is derived from the single-point Chern number~\cite{ceresoli_sp_prb_2007,favata_es_2023}, hence being very suited to large-scale \textit{ab initio} electronic structure simulations; that is relevant for the study of amorphous topological materials~\cite{corbae_EPL_2023,corbae_natmat_2023,fazzio_nanolett_2019}, quasi-crystals~\cite{huang_prl_2018,huang_prb_2018} and in presence of defects or interfaces. Crucially, our formula contains the operator $\mathbf{r}$ only as an exponent in $e^{i\mathbf{b}_{1,2}\cdot \mathbf{r}}$, which is known to be the correct approach to treat the position operator in PBCs~\cite{resta_prl_1998,aligia_prl_1999} and is related to the modern theory of polarization based on the Berry phase~\cite{resta_review_1994}. We note that the PBC position operator has already proved to be key in calculating topological invariants in real space~\cite{prodan_prl_2010}, also for interacting systems~\cite{aligia_PRB_2023,gilardoni_prb_2022}. We have also provided physical insights on the connection between the PBC single-point/local Chern invariants and the Bott index: measuring the non-commutativity of the projected position operators is essentially measuring the impossibility of choosing a smooth gauge all over the BZ in the limit of a large supercell with single-point sampling.  
Finally, we emphasize that our approach is potentially very general and could be applied to any geometrical and topological quantity of the electronic structure, in any dimension. The only requirement to develop similar PBC local markers is that the given quantity must be a ground-state bulk property of the system and admit a single-point formula in the large supercell limit.

We acknowledge useful discussions with R. Resta, A.M. acknowledges support from the ICSC --~Centro Nazionale di Ricerca in High Performance Computing, Big Data and Quantum Computing, funded by European Union --~NextGenerationEU --~PNRR, Missione 4 Componente 2 Investimento 1.4.

\bibliography{biblio}

\clearpage
\onecolumngrid

\renewcommand{\figurename}{Supplementary Fig.}
\renewcommand{\tablename}{Supplementary Tab.}
\renewcommand\thefigure{\arabic{figure}}
\renewcommand\thetable{\arabic{figure}}
\renewcommand\thesubsection{Supplementary Note \arabic{subsection}}
\renewcommand\theequation{S\arabic{equation}}
\setcounter{equation}{0}
\setcounter{figure}{0}
\setcounter{table}{0}
\setcounter{subsection}{0}
\title{Supplementary Material for \\ Local Chern Marker for Periodic Systems}
\date{\today}
\author{Nicolas Baù}
\affiliation{\trieste}
\author{Antimo Marrazzo}
\email{antimo.marrazzo@units.it}
\affiliation{\trieste}

\newcommand{\pg}{\mathcal P_{\Gamma}}
\newcommand{\com}[1]{\big[ #1 \big]}
\newcommand{\imtr}[1]{\mathrm{Im\,Tr}\Big\{ #1 \Big\}}
\newcommand{\imtrv}[1]{\mathrm{Im\,Tr_A}\Big\{ #1 \Big\}}

\maketitle
\section{Supplementary material}
\subsection{Local Chern marker for a finite system with periodic boundary conditions}

  In the limit of a finite sample centered at the origin of a much larger supercell with periodic boundary conditions, the wavefunction is non-vanishing only for coordinates that are always much smaller than the linear size of the system, so it holds that $r_j\ll L_j$. We recall that the dual states appearing in the definition of the PBC local Chern marker are defined as:
  \begin{eqnarray}
    \ket{\tilde u_{n\mathbf b_j}} = \sum_{m=1}^{N_{occ}}\big[S^{-1}(\mathbf b_j)\big]_{mn}e^{-i\mathbf b_j\cdot\mathbf r}\ket{u_{m\Gamma}}
  \end{eqnarray}
  with the overlap matrix $S_{mn}(\mathbf b_j)=\braket{u_{m\Gamma}|e^{-i\mathbf b_j\cdot\mathbf r}|u_{n\Gamma}}$.
  We can expand the exponentials appearing in the definition of the dual states, so that the projector $\mathcal P_{\mathbf b_j}$ ($j=1, 2$) can be written as
  \begin{align}
    \mathcal P_{\mathbf b_j}&=\sum_{n=1}^{N_{occ}}\ket{\tilde u_{n\mathbf b_j}}\bra{\tilde u_{n\mathbf b_j}}=\nonumber\\
    &=\sum_{n=1}^{N_{occ}}\sum_{\alpha=1}^{N_{occ}}\sum_{\beta=1}^{N_{occ}}\big[S^{-1}(\mathbf b_j)\big]_{\alpha n}e^{-i\mathbf b_j\cdot\mathbf r}\ket{u_{\alpha\Gamma}}\bra{u_{\beta\Gamma}}e^{i\mathbf b_j\cdot\mathbf r}\big[S^{-1}(\mathbf b_j)\big]_{\beta n}^*=\nonumber\\
    &=\sum_{n,\alpha,\beta=1}^{N_{occ}}\big[S^{-1}(\mathbf b_j)\big]_{\alpha n}\left[\mathbb I-i\mathbf b_j\cdot\mathbf r-\frac{1}{2}(\mathbf b_j\cdot \mathbf r)^2\right]\ket{u_{\alpha\Gamma}}\bra{u_{\beta\Gamma}}\left[\mathbb I+i\mathbf b_j\cdot\mathbf r-\frac{1}{2}(\mathbf b_j\cdot \mathbf r)^2\right]\big[S^{-1}(\mathbf b_j)\big]_{\beta n}^* + \mathcal{O}(L_j^{-3}).
  \end{align}
  With no loss of generality, we use a square supercell so that $\mathbf b_j\cdot\mathbf r=2\pi r_j/L_j$ where $r_1=x$ and $r_2=y$ are the Cartesian component of the position operator. We start our calculation by evaluating the term of order zero in $L_j$:
  \begin{align}
    &\sum_{n,\alpha,\beta=1}^{N_{occ}}\big[S^{-1}(\mathbf b_j)\big]_{\alpha n}\ket{u_{\alpha\Gamma}}\bra{u_{\beta\Gamma}}\big[S^{-1}(\mathbf b_j)\big]_{\beta n}^*=\sum_{\alpha,\beta=1}^{N_{occ}}\ket{u_{\alpha\Gamma}}\bra{u_{\beta\Gamma}}\sum_{n=1}^{N_{occ}}\big[S^{-1}(\mathbf b_j)\big]_{\alpha n}\big[S^{-1}(\mathbf b_j)\big]_{\beta n}^*=\nonumber\\
    &=\sum_{\alpha,\beta=1}^{N_{occ}}\ket{u_{\alpha\Gamma}}\bra{u_{\beta\Gamma}}\sum_{n=1}^{N_{occ}}\big[S^{-1}(\mathbf b_j)\big]_{\alpha n}\big\{\big[S^{\dagger}(\mathbf b_j)\big]^{-1}\big\}_{n \beta} = \sum_{\alpha,\beta=1}^{N_{occ}}\ket{u_{\alpha\Gamma}}\bra{u_{\beta\Gamma}}\big[S^{\dagger}(\mathbf b_j)S(\mathbf b_j)\big]^{-1}_{\alpha\beta}.\label{eq:zeroorder}
  \end{align}
  In particular, by definition of the overlap matrix $S(\mathbf b_j)$, it holds that
  \begin{align}
    &\big[S^{\dagger}(\mathbf b_j)S(\mathbf b_j)\big]_{\alpha\beta}=\sum_{n=1}^{N_{occ}}\big(\braket{u_{\alpha\Gamma}|e^{-2\pi i r_j/L_j}|u_{n\Gamma}}\big)^{\dagger}\braket{u_{n\Gamma}|e^{-2\pi i r_j/L_j}|u_{\beta\Gamma}}=\nonumber\\
    &=\sum_{n=1}^{N_{occ}}\braket{u_{\alpha\Gamma}|e^{2\pi i r_j/L_j}|u_{n\Gamma}}\braket{u_{n\Gamma}|e^{-2\pi i r_j/L_j}|u_{\beta\Gamma}}=\braket{u_{\alpha\Gamma}|e^{2\pi i r_j /L_j}\mathcal P_{\Gamma}e^{-2\pi i r_j/L_j}|u_{\beta\Gamma}}
  \end{align}
  where $\mathcal P_{\Gamma}$ is the ground state projector. Now, applying the expansion $e^{iS}Oe^{-iS}=O+i[S,O]+\frac{i^2}{2!}[S,[S,O]]+\dots$ (see e.g. Ref.~\cite{Fetter}), we can rewrite:
  \begin{eqnarray}
    \big[S^{\dagger}(\mathbf b_j)S(\mathbf b_j)\big]_{\alpha\beta}=\delta_{\alpha\beta}+\frac{2\pi i}{L_j}\bra{u_{\alpha\Gamma}}\big[r_j,\mathcal P_{\Gamma}\big]\ket{u_{\beta\Gamma}}-\frac{2\pi^2}{L_j^2}\bra{u_{\alpha\Gamma}}\big[r_j,\big[r_j,\mathcal P_{\Gamma}\big]\big]\ket{u_{\beta\Gamma}} + \mathcal{O}(L_j^{-3}).
  \end{eqnarray}
  In order to evaluate the inverse matrix in the limit of a large supercell, we can expand it as a power series:
  \begin{align}
    \big[S^{\dagger}(\mathbf b_j)&S(\mathbf b_j)\big]_{\alpha\beta}^{-1}=\delta_{\alpha\beta}-\Big\{\frac{2\pi i}{L_j}\bra{u_{\alpha\Gamma}}\big[r_j,\mathcal P_{\Gamma}\big]\ket{u_{\beta\Gamma}}-\frac{2\pi^2}{L_j^2}\bra{u_{\alpha\Gamma}}\big[r_j,\big[r_j,\mathcal P_{\Gamma}\big]\big]\ket{u_{\beta\Gamma}}\Big\} +\nonumber\\&\hspace{1cm} + \Big\{\frac{2\pi i}{L_j}\bra{u_{\alpha\Gamma}}\big[r_j,\mathcal P_{\Gamma}\big]\ket{u_{\beta\Gamma}}-\frac{2\pi^2}{L_j^2}\bra{u_{\alpha\Gamma}}\big[r_j,\big[r_j,\mathcal P_{\Gamma}\big]\big]\ket{u_{\beta\Gamma}}\Big\} ^2 + \mathcal{O}(L_j^{-3})=\\
    &=\delta_{\alpha\beta}-\frac{2\pi i}{L_j}\bra{u_{\alpha\Gamma}}\big[r_j,\mathcal P_{\Gamma}\big]\ket{u_{\beta\Gamma}}+\frac{2\pi^2}{L_j^2}\bra{u_{\alpha\Gamma}}\big[r_j,\big[r_j,\mathcal P_{\Gamma}\big]\big]\ket{u_{\beta\Gamma}}-\nonumber\\&\hspace{1cm}-\frac{4\pi^2}{L_j^2}\bra{u_{\alpha\Gamma}}\big[r_j,\mathcal P_{\Gamma}\big]^2\ket{u_{\beta\Gamma}}+\mathcal{O}(L_j^{-3}).\label{eq:sdags}
  \end{align}
  Now, substituting Eq.~\ref{eq:sdags} into Eq.~\ref{eq:zeroorder}, the term of order zero in the expansion becomes:
  \begin{align}
    \sum_{\alpha,\beta=1}^{N_{occ}}&\ket{u_{\alpha\Gamma}}\bra{u_{\beta\Gamma}}\big[S^{\dagger}(\mathbf b_j)S(\mathbf b_j)\big]^{-1}_{\alpha\beta} + \mathcal{O}(L_j^{-3})=\nonumber\\ &=\pg-\frac{2\pi i}{L_j}\pg\com{r_j,\pg}\pg + \frac{2\pi^2}{L_j^2}\pg\com{r_j, \com{r_j,\pg}}-\frac{4\pi^2}{L_j^2}\pg\com{r_j,\pg}\com{r_j,\pg}\pg + \mathcal{O}(L_j^{-3})=\nonumber\\
    &=\pg+\mathcal{O}(L_j^{-3})\label{eq:zeroterm}
  \end{align}
  by evaluating explicitly the commutators. Following the same logic, the first-order terms of the expansion for the exponentials yield:
  \begin{align}
    &a) & -\frac{2\pi i}{L_j}\sum_{\alpha,\beta=1}^{N_{occ}}&r_j\ket{u_{\alpha\Gamma}}\bra{u_{\beta\Gamma}}\big[S^{\dagger}(\mathbf b_j)S(\mathbf b_j)\big]^{-1}_{\alpha\beta}+\mathcal{O}(L_j^{-3})=\nonumber\\
    &&&=-\frac{2\pi i}{L_j}\Big\{ r_j\pg-\frac{2\pi i}{L_j}r_j\pg\com{r_j,\pg}\pg \Big\}+\mathcal{O}(L_j^{-3})=-\frac{2\pi i}{L_j}r_j\pg+\mathcal{O}(L_j^{-3})\hspace{1cm} \label{eq:oneterma}\\
    &b) & \frac{2\pi i}{L_j}\sum_{\alpha,\beta=1}^{N_{occ}}&\ket{u_{\alpha\Gamma}}\bra{u_{\beta\Gamma}}r_j\big[S^{\dagger}(\mathbf b_j)S(\mathbf b_j)\big]^{-1}_{\alpha\beta}+\mathcal{O}(L_j^{-3})=\frac{2\pi i}{L_j}\pg r_j + \mathcal{O}(L_j^{-3}).\label{eq:onetermb}
  \end{align}
  Considering the second order terms of the expansion we have:
  \begin{align}
    &a) & -\frac{2\pi^2}{L_j^2}\sum_{\alpha,\beta=1}^{N_{occ}}&r_j^2\ket{u_{\alpha\Gamma}}\bra{u_{\beta\Gamma}}\big[S^{\dagger}(\mathbf b_j)S(\mathbf b_j)\big]^{-1}_{\alpha\beta}+\mathcal{O}(L_j^{-3})=-\frac{2\pi^2}{L_j^2}r_j^2\pg \hspace{3.5cm}\label{eq:twoterma}\\
    &b) & \frac{2\pi^2}{L_j^2}\sum_{\alpha,\beta=1}^{N_{occ}}&\ket{u_{\alpha\Gamma}}\bra{u_{\beta\Gamma}}r_j^2\big[S^{\dagger}(\mathbf b_j)S(\mathbf b_j)\big]^{-1}_{\alpha\beta}+\mathcal{O}(L_j^{-3})=\frac{2\pi^2}{L_j^2}\pg r_j^2+\mathcal{O}(L_j^{-3}) \label{eq:twotermb}\\
    &c) & \frac{4\pi^2}{L_j^2}\sum_{\alpha,\beta=1}^{N_{occ}}&r_j\ket{u_{\alpha\Gamma}}\bra{u_{\beta\Gamma}}r_j\big[S^{\dagger}(\mathbf b_j)S(\mathbf b_j)\big]^{-1}_{\alpha\beta}+\mathcal{O}(L_j^{-3})=\frac{4\pi^2}{L_j^2}r_j\pg r_j+\mathcal{O}(L_j^{-3}).\label{eq:twotermc}
  \end{align}
  Finally, summing the contribution of Eqs.~\ref{eq:zeroterm}-\ref{eq:twotermc} we arrive at the expansion of the projectors $\mathcal P_{\mathbf b_j}$ in the limit of a large supercell up to order $L_j^{-3}$:
  \begin{eqnarray}\label{eq:pbj}
    \mathcal P_{\mathbf b_j}=\pg -\frac{2\pi i}{L_j}\com{r_j,\pg}-\frac{2\pi^2}{L_j^2}\com{r_j,\com{r_j,\pg}}+\mathcal{O}(L_j^{-3}).
  \end{eqnarray}
  We are now ready to calculate the expansion of the PBC local Chern marker: as discussed in Ref.\cite{bianco_prb_2011} it is crucially to address the commutators first and take the trace only at the end. Hence, we start by considering the expression of the single-point Chern number expressed as total trace (see main text):
  \begin{eqnarray}\label{eq:pbclcm}
   C=-\frac{1}{2\pi}\imtr{\com{\mathcal P_{\mathbf b_1},\mathcal P_{\mathbf b_2}}\pg}.
  \end{eqnarray}
  To calculate its expansion in the large supercell limit, we use the expansion of Eq.~\ref{eq:pbj} in Eq.~\ref{eq:pbclcm} and ignore the terms of order $L^{-3}$ for both the $x$ and $y$ direction:
  \begin{eqnarray}
   C&=-\dfrac{1}{2\pi}\imtr{\com{\pg -\dfrac{2\pi i}{L_x}\com{x,\pg}-\dfrac{2\pi^2}{L_x^2}\com{x,\com{x,\pg}},\pg -\dfrac{2\pi i}{L_y}\com{y,\pg}-\dfrac{2\pi^2}{L_y^2}\com{y,\com{y,\pg}}}\pg} + \mathcal{O}(L^{-3}).
  \end{eqnarray}
  Exploiting the linearity of the commutator we see that:
  \begin{itemize}
    \item The terms containing $\pg$ do not contribute since $\com{\pg,\pg}=0$;
    \item The terms of order $L^{-1}$ do not contribute since we can apply the cyclic property of the trace and see that:
    \begin{eqnarray}\label{eq:del1}
      \imtr{\com{\com{x,\pg},\pg}\pg} = \imtr{x\pg-\pg x\pg}=0
    \end{eqnarray}
    and the same holds for the contribution comprising the $y$ operator;
    \item The terms containing $r_j^2$ do not contribute since applying the cyclic property of the trace:
    \begin{eqnarray}\label{eq:del2}
      \imtr{\com{\com{x,\com{x,\pg}},\pg}\pg}=\imtr{x^2\pg-\pg x^2+2\pg x\pg x-2x\pg x\pg}=0
    \end{eqnarray}
    and the same holds for the contribution comprising the $y$ operator;
  \end{itemize}
  From Eqs.~\ref{eq:del1},\ref{eq:del2} we see that the only contribution left is
  \begin{align}
    C&=-\dfrac{1}{2\pi}\left(-\frac{4\pi^2}{L_xL_y}\right)\imtr{\com{\com{x,\pg},\com{y,\pg}}\pg} + \mathcal{O}(L^{-3})=\\&=\frac{2\pi}{A_{sc}}\imtr{\com{\com{x,\pg},\com{y,\pg}}\pg} + \mathcal{O}(L^{-3})
  \end{align}
  where $A_{sc}=L_xL_y$ is the area of the supercell.
  We can now evaluate first the commutator and only later on taking the trace, as suggested in Ref.~\cite{bianco_prb_2011}:
  \begin{align}
    C&=2\pi\imtrv{\com{x,\pg}\com{y,\pg}\pg-\com{y,\pg}\com{x,\pg}\pg} + \mathcal{O}(L^{-3})=\nonumber\\
    &=2\pi\imtrv{\pg x\pg y\pg-\pg y\pg x\pg} + \mathcal{O}(L^{-3})=2\pi\imtrv{\com{\pg x\pg, \pg y \pg}} + \mathcal{O}(L^{-3})
  \end{align}
  Finally, exploiting the relation $\imtr{\pg x\pg y}=\frac{1}{2i}\mathrm{Tr}\Big\{ \com{\pg x \pg, \pg y \pg} \Big\}$, we arrive at the OBC definition of the local Chern marker introduced in Ref.~\cite{bianco_prb_2011}:
  \begin{eqnarray}
    \label{eq:final_sm}
    \lim_{x\ll L_x,y\ll L_y}\mathcal C(\mathbf r)=4\pi\braket{\mathbf{r}|\pg x\pg y|\mathbf{r}}=-4\pi\braket{\mathbf{r}|\pg x (\mathbb I-\pg)y|\mathbf{r}} = \mathfrak{C}(\mathbf r)
  \end{eqnarray}
The space-resolved Chern number can be then calculated as the local trace per unit area (macroscopic average) of the Chern marker defined by Eq.~\ref{eq:final_sm}.

\end{document}